\documentclass[epj,final]{svjour}

\usepackage{latexsym}
\usepackage{url}
\usepackage{amsfonts}
\usepackage{amsmath, amssymb}
\RequirePackage{graphicx}

\begin{document}
\title{Ultrahigh accuracy time synchronization technique operation on the Moon}
\author{V.G.~Gurzadyan
, A.T. Margaryan 
}                     
%
%
\institute{Center for Cosmology and Astrophysics, Alikhanian National Laboratory and Yerevan State University, Yerevan, Armenia}
\date{Received: date / Revised version: date}
%

\abstract{Ultrahigh accuracy time synchronization technique based on the Optical Frequency Comb and the GHZ radiofrequency spiral scanning deflector is suggested to install on the Moon during the ARTEMIS mission planned by NASA for 2024.  The comparison with the parameters of an analogous device operated in the Earth's gravity will enable the testing to high accuracy fundamental physical principles.
} 
\PACS{
      {95.30.Sf}{Relativity and gravitation}   
} 
\maketitle

\section{Introduction}

The installation of retroreflectors on the Moon by Apollo 11 and by subsequent missions and Lunar laser ranging opened entire new paths for research, from testing of fundamental physics, General Relativity, Nordtvedt effect\footnote{Ken Nordtvedt had told to one of the authors (VG), how he approached Robert Dicke, the then NASA official, during a joint air flight, suggesting to include a retroreflector in Apollo 11 mission.}, up to a bunch of issues of the structure and evolution of the Moon. The present proposal\footnote{This is the extended version of the Artemis Science White Paper N.2013, NASA.} aims the operation on the Moon during the ARTEMIS mission planned for 2024 of to date the highest accuracy synchronization technique.    

Present study is based on authors’ experience on to date the highest accuracy test of light speed invariance (Kennedy –- Thorndike test) at the GRAAL experiment at European Synchrotron Radiation Facility (ESRF, Grenoble) \cite{1}, General Relativity’s Lense-Thirring effect testing via LARES satellite laser ranging measurements \cite{2,3}.   

Femtosecond optical frequency combs (OFC) are revolutionizing precision measurements of time and frequency [4]. The simplicity, robustness and improved precision femtosecond lasers have now led to their prominence in the field of optical frequency metrology [5]. In addition, their use is developing significant new time domain applications based on the precise control of the carrier-envelope phase, as involved in the present proposal. 

\section{The RF timing technique}

The Radio Frequency (RF) timing technique converts the information in the time domain to a spatial, linear domain that allows measurement of time with such high precision (1 $ps$ or less) by mean of optical sensor, the RFPMT (Radio Frequency Photo Multiplier Tube) \cite{6,7,8,9}. No optical sensor is capable of matching the combination of ultra-high timing resolution and very fast readout speed promised by the RFPMT, making it ideal for Time Correlated Single Photon Counting (TCSPC) applications. Schematically the small area of photocathode RFPMT is presented in Figure 1. 

 \begin{figure}[h]
\caption{Schematic layout of the RFPMT. (1) glass-window, (2) photo-cathode, (3) accelerating electrode, (4) electrostatic lens, (5) RF deflector, (6) microchannel plate based electron detector, (7) position sensitive readout system.}
\centering
\includegraphics[]{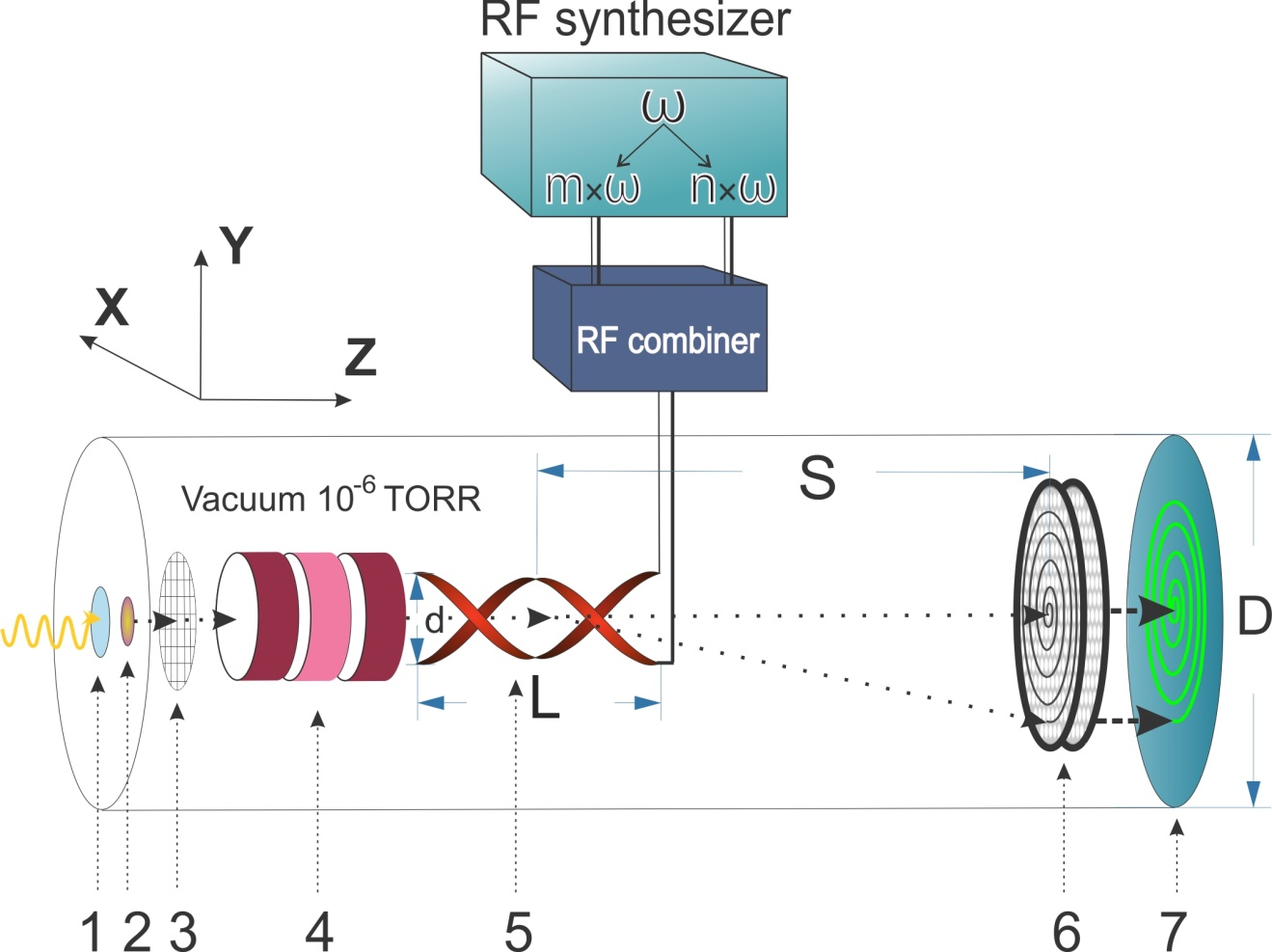}
\end{figure}

The primary photon pulse passes quartz glass widow (1) hits the photocathode (2) and produces photo-electrons (PEs). The produced PEs are accelerated by a voltage V (about 2.5 kV) applied between the photocathode and an electron transparent electrode (3), mounted at  about 1mm far from photocathode and enters into the electrostatic lens (4). The electrostatic lens then focuses and transposes the electrons isochronously onto the position sensitive PE detector (6,7) placed at the end of tube. Along the way the electrons pass through the circular sweep RF deflection system (5), consisting of dedicated electrodes and the RF power source.  The idea is similar to the Streak Camera principle and based on the conversion of time related to the single photon signal to a spatial one detected by the Detector (6,7) in Figure 1. The internal time resolution of such a system can reach a level of about 1 $ps$. The crucial and challenging aspects are 1) the RF deflector and 2) the digitization of the readout. 

1.	The RF deflector and the anode readout architecture: The sensitivity of the RF deflector at the dedicated resonance frequencies is about 0.1 radian/$W^{1/2}$ and about 20 $V$ (peak to peak) RF sine wave is sufficient to produce a scanning circle with a few cm radius on the PE detector plane.  Spiral scanning will increase the $ps$ period. Two helical deflectors at slightly different RF frequencies produce a “beat” in the amplitude of the scanned circle which sets the period for a single pico-time spiral sequence. 

2.	Readout and Digitization: Two readout methods have been devised to locate the position on the scanned circle: interpolation or 1D readout and pixel-by-pixel or 2D readout. Interpolation readout, using a phase sensitive delay line anode, needs only two readout channels and position is determined by applying delay-line time difference technique. However, it can only bear a moderately high counting rate. For example, with the dual MCP chevron assembly the anticipated maximum rate is about 1 MHz. A pixel-by-pixel readout anode will permit much higher counting rates. 

The combination of the spiral scanning deflector of keV electrons with the recently developed 55 $\mu m$ pixel based on TimePix3Cam will result in 1 $ps$ resolution, 1 THz bandwidth and about 100 MHz throughput rate single photon counting device. The about 1 $ps$ timing resolution of the RFPMT is a factor 10-100 better than conventional PMT or SPAD sensors. The RF timing approach obviously has advantages in comparison with the recently developed superconducting nanowire single-photon detectors. Coupled to a high rate capability with minimal dead time this will turn the RFPMT as an efficient photon detector.

Narrow linewidth lasers referenced to optical transitions in atoms and ions are the best electromagnetic references, with projected fractional frequency instability and uncertainties well below $10^{-15} \tau^{-1/2}$  ($\tau$ is the time) and $10^{-18}$, respectively. When used in conjunction with such ultraprecise frequency standards, the femtosecond laser serves as a broadband synthesizer that phase-coherently converts the input optical frequency to microwave frequencies. The excess fractional frequency noise introduced in the synthesis process can approach the level $10^{-19}$ \cite{5}. 

Combination of femtosecond optical frequency combs (OFC) with a radio frequency timing technique of visible photons \cite{6,7} provides this tool for precision time or frequency measurements \cite{8} based on the comparative operation of the optical and radio frequency techniques. The GHz radiofrequency spiral scanning deflector is capable for real-time conversion of a sequence of electrons into two-dimensional spatial images, synchronously with the optical frequency comb. This ultrafast timing processor creates a sequence of keV energy electrons directly encoded onto the spatial locus, producing a two-dimensional spatial image of the electrons arrival time. This produces a THz bandwidth and a dead-time free device for electron detection which is capable of achieving 1 $Tbit/s$ sampling rate and 1 $ps$ time resolution in a few hundred $ns$ time range. 

\section{Optoelectronic heterodyne}

The synchroscan mode of the RF phototube is used for precision comparison of the frequencies of two OFCs \cite{8,9} as shown in Figure 2. The photon beam from OFC 1 with a repetition rate $\nu_1^0$   is split in two parts. The first part is directed to the photodiode and RF synthesizer, which generates a sinusoidal signal $V_{RF}(t)$ synchronous to the photon beam 1 for driving the RF phototube
\begin{equation}
V_{RF}(t)=V_{RF}^0 \sin[2\pi \nu_1^0 t + \phi_{RF}(t) + \phi_{RF}^0].
\end{equation}

The second part of the photon beam illuminates the photocathode and the resulting photoelectrons are extracted, accelerated, focused, deflected by an RF deflector and detected on the scanning circle with a phase
\begin{equation}
\Phi_1(t)=\phi_1(t)+\phi_{1RF}^0=\phi_{1RF}(t) + \phi_{1T}(t) + \phi_{1RF}^0.
\end{equation}

The photoelectrons from the OFC 2 photon beam detected by the system will have a phase

\begin{equation}
\Phi_2(t) = 2\pi (\nu_1^0 - \nu_2^0)t + \phi_{2}(t) + \phi_{2RF}^0 =   2\pi (\nu_1^0 - \nu_2^0)t + \phi_{2RF}(t) + \phi_{2T}(t)   + \phi_{2RF}^0.
\end{equation}

\begin{figure}[h]
\caption{Functional diagram of the optoelectronic heterodyne. BS- beam splitter, M- reflector, PD- photo diode, 1- photocathode, 2- accelerating electrode, 3- electrostatic lens, 4- deflection electrode, 5- quarter wavelength coaxial cavity, 6- position sensitive secondary electron detector, 7- arbitrary reference fixed on SE detector readout system, 8- and 9- PE’s from Laser 1 and Laser 2. 
}
\centering
\includegraphics[width=15cm]{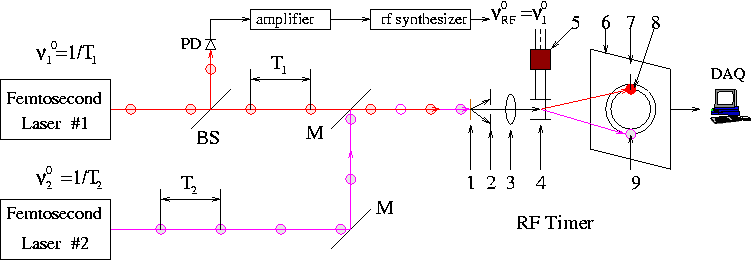}
\end{figure}

The two beam spots drift with a repetition rate  $\nu_H=\nu_1^0-\nu_2^0$  and the long term instability of this $\Phi_2(t) - \Phi_1(t)= 2\pi (\nu_1^0-\nu_2^0)$  beat signal is about or less than 200 $fs/hr$ \cite{10}. Therefore, the optoelectronic heterodyning system based on the RF phototube and in conjunction with optical clocks will provide precisions comparable to those of optical clocks. This technique is useful for measuring tiny beat frequency of two practically equal frequencies $\nu_1^0$ and $\nu_2^0$. In the case $\nu_1^0 = \nu_2^0$ the difference between frequencies can be determined with precision equal to
\begin{equation} 
\frac{\nu_1^0}{\nu_2^0} -1 = \sqrt 2 \, 200 \, fs/\tau,
\end{equation}

where $\tau$  is an averaging time interval. For $\tau=10^5 s$,  $\frac{\nu_1^0}{\nu_2^0} -1$ can reach the value $3\, 10^{-18}$, if the instabilities of femtosecond lasers are less than  $3\, 10^{-18}$.

\section{Gravitational redshift, equivalence principle}

A typical gravitational redshift experiment measures the frequency or wavelength shift  $Z\equiv \Delta \nu/\nu$ between two identical frequency standards (clocks) placed at rest in different static gravitational fields. As a consequence of the Einstein Equivalence Principle (EEP) \cite{11,12,13} under the assumption that the Weak Equivalence Principle (WEP) and Local Lorentz Invariance (LLI) gravitational redshift can be used as direct probe for Local Position Invariance (LPI) by comparing two clocks in different gravitational potentials. In weak-field limit a parameter $\alpha^A$  is linked to the difference of the Newtonian gravitational potential $\Delta U$  between the locations of the clocks as
\begin{equation}
Z= ( 1 + \alpha^A) \frac{\Delta U}{c^2}.
\end{equation}

LPI violation was traced via Cs, H, Mg clocks, cavity oscillators, optical clocks. Limits on  can be obtained comparing two identical clocks at different potentials or by a null redshift experiment, comparing non-identical co-located clocks A and B in a time-varying gravitational potential i.e.
\begin{equation}
\frac{\nu^A}{\nu^B} =  (\frac{\nu^A}{\nu^B})_0 ( 1+  (\alpha^A - \alpha^B)) \frac{\Delta U}{c^2},
\end{equation}
where $(\nu^A/\nu^B)$  is the locally measured frequency ratio and $(\nu^A/\nu^B)_0$ is that ratio at some initial location. The bound from comparing a Cesium atomic fountain with a Hydrogen maser for a year yields $\alpha^H - \alpha^{Cs}  < 2.1 \, 10^{-5}$ \cite{12,13}. The limit obtained at the Gravity Probe-A experiment by means of a rocket-borne H-maser was $\alpha < 7 \, 10^{-5}$  \cite{14,15}. 

The experimental arrangement for identical clocks on the Moon and the Earth is shown schematically in Fig. 3. Both optical clocks are referenced to the same atomic transition and provide two OFCs and precise comparison of the time signals on each clock located at the Moon and the Earth, including associated relevant effects (e.g. \cite{16}).  The same atomic transition in weak-field metric for the Moon and the Earth, 
\begin{equation}
g_{00} = - (1 + 2\varphi),
\end{equation}
yields $\nu_{Earth}/\nu_{Moon} -1 = \varphi_{Earth} - \varphi_{Moon}$, where $\varphi_{Moon}$ and $\varphi_{Earth}$ are the gravitational potentials, respectively. The beat signal $\nu_{Earth} - \nu_{Moon}$ will cause counterclockwise and clockwise drifts of beam spots relative to each other, on the scanning circles of the A and B RF timers, consequently. The speed of these drifts is about 0.66 ns/s and amount to 6600 ns in a $10^4$ s averaging time. As the stability of the proposed technique is better than 200 $fs/hr$, the gravitational red shift difference of the Earth and Moon could be measured and monitored with a precision about or better than $3 \, 10^{-8}$.  
 
\begin{figure}[h]
\caption{Schematic of the gravitational red-shift experiment based on optoelectronic heterodyne. BS- beam splitter, M- reflector, PD- photo diode, 1- photocathode, 2- accelerating electrode, 3- electrostatic lens, 4- deflection electrode, 5- quarter wavelength coaxial cavity, 6- position sensitive secondary electron detector, 7- arbitrary reference fixed on SE detector readout system, 8 and 9- accelerated and deflected photoelectrons directly from 1 and 2 optical clocks located on the Moon and Earth.}
\centering
\includegraphics[width=15cm]{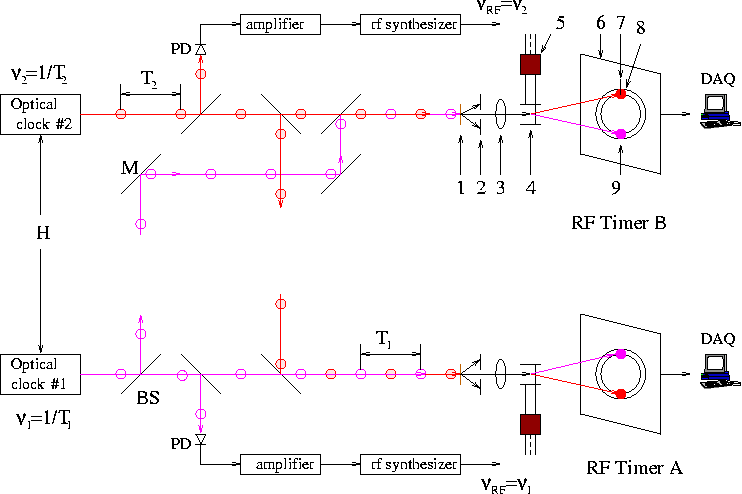}
\end{figure}

\section{Conclusions}

Femtosecond optical frequency combs in combination with a radio-frequency timing technique results in a highly stable (<200 fs/hr) optoelectronic heterodyne are suggested to apply for Moon-Earth gravitational redshift precise experiments during the Artemis mission. The use of radiofrequency technique will provide precisions of 1 $ps$ for single photons with fast readouts and essentially no dead-time. The stability of the technique is better than 200 fs/hr, therefore the gravitational redshift difference of the Earth and Moon could be measured and monitored with a precision about or better than $3\, 10^{-8}$. The GHz radiofrequency timing technique with such ultrahigh level of precision in time measurements operated in the Moon gravity, then compared with the analogous device operation in the Earth’s gravity, will enable the testing to high accuracy fundamental physical principles. 

Various technical realization options and solutions can be considered to reach the possibly lightest, spatially smallest transportable device. 

This work is partially supported by International Science and Technology Center (ISTC) grant A-2390.



\end{document}